\documentclass[letterpaper, 10 pt, conference]{ieeeconf}  

\IEEEoverridecommandlockouts                              

\usepackage{graphics} 
\usepackage{epsfig} 
\usepackage{mathptmx} 
\usepackage{times} 
\usepackage{amsmath} 
\usepackage{amssymb}  
\usepackage{psfrag} 
\usepackage{float}
\usepackage{xcolor}

\title{\LARGE \bf
On the Design of Limit Cycles of Planar Switching Affine Systems
}

\author{Nils Hanke$^{1}$ and Olaf Stursberg$^{2}$ 
\thanks{$^{1}$Nils Hanke, Control and System Theory, Dept. of Electrical Eng. and Computer Science,  University of Kassel, Germany
        {\tt\small n.hanke@uni-kassel.de}}%
\thanks{$^{2}$Olaf Stursberg, Control and System Theory, Dept. of Electrical Eng. and Computer Science, University of Kassel, Germany
        {\tt\small stursberg@uni-kassel.de}}%
}

\newcommand{\RN}[1]{\uppercase\expandafter{\romannumeral#1}}
\newtheorem{theorem}{Theorem}

\begin{document}

\maketitle
\thispagestyle{empty}
\pagestyle{plain} 

\begin{abstract}
In the context of studying periodic processes, this paper investigates first under which conditions switching affine systems in the plane generate stable limit cycles. Based on these conditions, a design methodology is proposed by which the phase portraits of the switching systems are determined to obtain globally stable limit cycles from simple specifications, such as given amplitudes and frequencies of desired oscillations. As an application, the paper finally shows that an oscillator model can be derived with a small effort from data measured for an unknown oscillating system.
\end{abstract}

\section{INTRODUCTION}
Since oscillations occur in a large variety of domains (technical, physical, biological, astronomical, etc.), limit cycles have been a subject of research for a long period of time, see e.g. \cite{OHNO.2006, Teplinsky.2008, Mirollo.1990,Peterchev.2003}.
They have been characterized and investigated for different nonlinear models such as Kuramoto, Van-der-Pol, or FitzHugh-Nagumo oscillators \cite{Kuramoto.1975,Joshi.2016,Gaiko.2011,Dorfler.2014}. 
The analytical characterization of limit cycles as well as the specification of conditions under which unique limit cycles are observed is, however, limited to very special cases. One thread of research has thus studied the existence of limit cycles for systems consisting of multiple linear systems: Goncalves considered limit cycles induced by feedback through relays with hysteresis for stable linear time-invariant systems, and he 
formulated conditions as linear matrix inequalities (LMIs) that guarantee global asymptotic stability of limit cycles  \cite{Goncalves.2001}. Follow-up investigations led to necessary conditions for regions of stability using piece-wise linear systems (PLS) \cite{Goncalves.2003}. The different approach in \cite{Arena.2019} aims at controlling the dynamics of networked piecewise-linear-shaped Fitz-Hugh-Nagumo neurons through their nullclines, relying on computing the oscillation period of all nodes in a master-slave configuration. Although PLS have been widely used in studies of nonlinear dynamical systems, the rigorous mathematical definition has been barely considered \cite{DiLaurea.2008}, \cite{Ponce.2015}. Alternative work has investigated hybrid control strategies which combine stabilizing and destabilizing control laws to obtain the properties of limit cycles with defined amplitudes and frequencies \cite{Schuldt.2014}. The occurrence of a stable limit cycle for a hybrid system in which the switching between different modes is determined by a suitable control strategy was described in \cite{Rubensson.1998}.Kai and Masuda designed a stable limit cycle as polygonal closed curve by connecting vertices of the polygon through line segments determined by piecewise affine systems \cite{Kai.2012}. Kai also provided conditions and analytic solutions to control the piecewise affine system by state feedback such that a given polygonal closed curve becomes a stable limit cycle \cite{Kai.2018,Kai.2019}.

In contrast to the previously cited work, the present paper follows the objective to design oscillating systems in the plane such that globally stable limit cycles are obtained. An important design requirement is that simple specifications (basically only the desired frequency and amplitudes) are sufficient to synthesize switching piece-wise systems, which e.g. can serve as a building block for networks of oscillators. The class of planar switching affine systems (PSAS) seems particularly suited for this purpose, as linear system theory allows representing limit cycles very efficiently, as opposed to the above-named examples of nonlinear systems with periodic behavior.
Accordingly, the paper first formulates a theorem for synthesizing globally stable limit cycles requiring just one switching surface. Based on an analytic specification of the limit cycle including the amplitude and frequency, a design approach is proposed that determines the phase portraits such that the specifications for amplitude and frequency are met, and stability is obtained. Finally the paper describes based on an example, how the design procedure can be used to quickly construct the periodic system to represent oscillations as found in experimental data, but without requiring techniques of system identification or machine learning.

\section{LIMIT CYCLE OF SWITCHING AFFINE SYSTEMS IN THE PLANE}
\newtheorem{Def}{Definition}
The underlying model class of this investigation is switching affine systems defined as follows: Assume a state space $X\subseteq\mathbb{R}^n$ with a polyhedral partition $\mathcal{P}=\{P^1,\ldots,P^{n_p}\}$ consisting of fully dimensional polyhedra which are pairwise disjoint with respect to their interior and which cover $X$. Let an affine autonomous dynamics:
\begin{equation}\label{SASmodel}
    \dot{x}(t)=A^i\cdot x(t)+b^i
\end{equation}
be assigned to any polyhedron $P^i\in\mathcal{P}$, $i\in\{1,\ldots,n_p\}$.
(Note that $A^i$ and $b^i$ may have been obtained from designing an affine state feedback controller $u(t)=K^i\cdot x(t)+ d^i$ for a plant model $\dot{x}(t)=\Tilde{A}^i\cdot x(t)+\Tilde{B}^i\cdot u(t)$ on $P^i$.) Let $T_k=\{t_0,t_1,\ldots,t_k,\}$ denote a (possibly infinite) set of switching times (extended by the initial time $t_0=0$).
A run of \eqref{SASmodel}, denoted by $\bar{x}_{[0,\infty[}$, starting from an initialization $x(t_0)=x_0$ is a sequence of phases $[t_k,t_{k+1}]$ in between two successive switching times, where the instance of \eqref{SASmodel} is activated for that $i$ for which $x(t)\in P^i$ for $t\in[t_k,t_{k+1}]$. For any \emph{switching} between two dynamics with indices $i,j\in\{1,\ldots n_p\}$ occurring at a time $t_{k}$ and for a state  $x(t_k)$, positioned on the shared boundary of $P^i$ and $P^{j}$, assume the following: The left-hand limit  of \eqref{SASmodel} in time 
$\dot{x}(t_k^-):=\lim_{\epsilon\rightarrow 0} A^i\cdot x(t-\epsilon)+b^i$ points outside of $P^i$, while the right-hand limit $\dot{x}(t_k^+):=\lim_{\epsilon\rightarrow 0} A^{j}\cdot x(t+\epsilon)+b^{j}$ points into $P^{j}$. This implies a unique switching time when the run crosses the boundary of $P^i$ and $P^j$. Note, however, that the destination $j$ of switching may not be unique, if $x(t_k)$ is contained in the boundaries of more than two polyhedra in $\mathcal{P}$. For this case, as well as for the initialization of $x(t_0)$ to a point on the shared boundary, an additional rule for selecting the active dynamics needs to be provided.  

Given the motivation of defining oscillations by a model, which is as simple as possible (but, of course, allows for stable limit cycles), the following considerations refer now to states defined in $\mathbb{R}^2$, and to switching affine systems with just $\mathcal{P}=\{P^1,P^2\}$. Obviously, the underlying partitioning then reduces to two half-spaces separated by a line, referred to as \emph{switching line} below. The corresponding model is defined as follows (using $\RN{1}$ and $\RN{2}$ to denote the two modes):\vspace{1mm}

\begin{Def}
\label{Definition1}{PSAS denoted as $\Sigma$}

Given a switching line $C\cdot x=d$ for $x\in\mathbb{R}^2$, $C=\lbrack c_{1,1},c_{1,2} \rbrack \in\mathbb{R}^{1\times 2}$, $d \in\mathbb{R}$, the polyhedral partition results to $\mathcal{P}=\{P^{\RN{1}},P^{\RN{2}}\}$ with 
$P^{\RN{1}}=\lbrace x| C\cdot x\leq d\rbrace$, $P^{\RN{2}}=\lbrace x |C\cdot x\geq d\rbrace$. For matrices $A^{\RN{1}}\in\mathbb{R}^{2\times 2}$, $b^{\RN{1}}\in\mathbb{R}^{2\times 1}$, $A^{\RN{2}}\in\mathbb{R}^{2\times 2}$, and $b^{\RN{2}}\in\mathbb{R}^{2\times 1}$, the dynamics assigned to this partition with $t\in\mathbb{R}_{\geq0}$ is: 
\begin{align}
\dot{x}(t)&=A^{\RN{1}}\cdot x(t)+b^{\RN{1}}\label{EQDynI}
\end{align}
for $x(t)\in Int(P^{\RN{1}})$, and:
\begin{align}
\dot{x}(t)&=A^{\RN{2}}\cdot x(t)+b^{\RN{2}}\label{EQDynII}
\end{align}
for $x(t)\in Int(P^{\RN{2}})$, where $Int$ denotes the interior of the respective set. For points on the switching line $C\cdot x(t)=d$, the dynamics \eqref{EQDynI} is assigned if $\lim_{\epsilon\rightarrow 0}
x(t-\epsilon)\in Int(P^{\RN{1}})$ applies for the predecessor in time,
and the dynamics \eqref{EQDynII} is assigned if $\lim_{\epsilon\rightarrow 0} x(t-\epsilon)\in Int(P^{\RN{2}})$. For the initial time $t_0$, $\Sigma$ starts to evolve according to \eqref{EQDynI}, if $x(t_0)\in Int(P^{\RN{1}})$, and according to \eqref{EQDynI} if $x(t_0)\in Int(P^{\RN{2}})$. For $C\cdot x(t_0)=d$, it is assigned by convention that $\Sigma$ starts to evolve with \eqref{EQDynII} if and only if $C\cdot (A^{\RN{2}}\cdot x(t)+b^{\RN{2}})>0$ and ( $C\cdot (A^{\RN{1}}\cdot x(t)+b^{\RN{1}})\geq 0$ or $\|C\cdot (A^{\RN{2}}\cdot x(t)+b^{\RN{2}})\|_2 > \|C\cdot (A^{\RN{1}}\cdot x(t)+b^{\RN{1}})\|_2$ ).
\hfill $\Box$  
\end{Def}\vspace{1mm}

While a run $\bar{x}_{[0,\infty[}$ of $\Sigma$ follows in general from the rules indicated below \eqref{SASmodel} and in Def.~1, the specific instance of a limit cycle is defined next: 

\begin{Def}
\label{Definition2}{Limit Cycle of $\Sigma$}

A run $\bar{x}^{*}_{\lbrack0,\infty\lbrack}$ of $\Sigma$ according to Def.~\ref{Definition1} is called \emph{limit cycle}, if a finite period $T\in\mathbb{R}_{>0}$ exists
such that for any point $x(t)\in \bar{x}^{*}_{\lbrack0,\infty\lbrack}$, $t\in\mathbb{R}_{\geq 0}$ it applies that: $x(t+T)=x(t)$. 
\hfill $\Box$  
\end{Def}\vspace{1mm}

The following theorem states conditions which ensure that the run of $\Sigma$ follows the limit cycle forever, if the run starts from a point on the cycle.

\begin{theorem}
\label{Theorem1}
For a system $\Sigma$ as specified in Def.~\ref{Definition1}, let the parametrization satisfy that $A^{\RN{1}}$, $A^{\RN{2}}$ have only distinct negative real eigenvalues $\lambda^{\RN{1}}=\lbrack \lambda^{\RN{1}}_{1,1},\lambda^{\RN{1}}_{2,1} \rbrack^T\in\mathbb{R}^{2x1}$, $\lambda^{\RN{2}}=\lbrack \lambda^{\RN{2}}_{1,2},\lambda^{\RN{2}}_{2,2} \rbrack^T\in\mathbb{R}^{2x1}$, and let unique equilibrium points $x^{\RN{1}}_{R}\in\mathbb{R}^{2\times 1}$, $x^{\RN{2}}_R\in\mathbb{R}^{2\times 1}$ follow from the choice of $A_i$. 
Then, a unique limit cycle $\bar{x}^{*}_{\lbrack0,\infty\lbrack}$ with period $T$ according to Def.~2 exists with initialization to 
$x(t_{s_0})$ with $C\cdot x(t_{s_0})=d$ and for two different switching points $x(t_{s_1})\neq x(t_{s_2})$ on the switching line, if the following set of sufficient conditions holds: 
\begin{align}
&C\cdot (A^{\RN{1}}\cdot x(t_{s_0})+b^{\RN{1}})<0  \label{EQ8}\\
&C\cdot (A^{\RN{1}}\cdot x(t_{s_1})+b^{\RN{1}})>0,\ C\cdot (A^{\RN{2}}\cdot x(t_{s_1})+b^{\RN{2}})>0\label{EQ9}\\
&C\cdot (A^{\RN{2}}\cdot x(t_{s_2})+b^{\RN{2}})<0\label{EQ11}\\
&C\cdot x^{\RN{1}}_R>d,\ C\cdot x^{\RN{2}}_R<d\label{EQ10}\\
&x(t_{s_2})=x(t_{s_0}), T=t_{s_2}-t_{s_0}.\label{EQ14}
\end{align}\hfill$\Box$
\end{theorem}

\begin{figure}
    \centering
   \psfrag{b2}[][r]{$\dot{x}^{\RN{1}}(t_{s_0})$}
    \psfrag{b1}[][r]{$\dot{x}^{\RN{1}}(t_{s_1})$}
    \psfrag{g2}[][r]{$x^{\RN{2}}_R$}
    \psfrag{c1}[][r]{$\dot{x}^{\RN{2}}(t_{s_1})$}
    \psfrag{c2}[][]{$\dot{x}^{\RN{2}}(t_{s_2})$}
    \psfrag{g1}[][]{$x^{\RN{1}}_R$}
    \psfrag{d2}[][r]{\quad $C\cdot x(t_{s_0})=C\cdot x(t_{s_2})=d$}
    \psfrag{x1}[][]{$P^{\RN{1}}$}
    \psfrag{x2}[][]{$P^{\RN{2}}$}
    \psfrag{d1}[][r]{$C\cdot x(t_{s_1})=d$}
    \psfrag{a}[][r]{$C$}
    \psfrag{h1}[][]{$\bar{x}^*_{[0,\infty[}$}
   \psfrag{h2}[][]{}
    \psfrag{i1}[][]{$x_1$}
    \psfrag{i2}[][]{$x_2$}
    \psfrag{j1}[][]{$\mathcal{A}_{*}$}
    \includegraphics[width=0.45\textwidth]{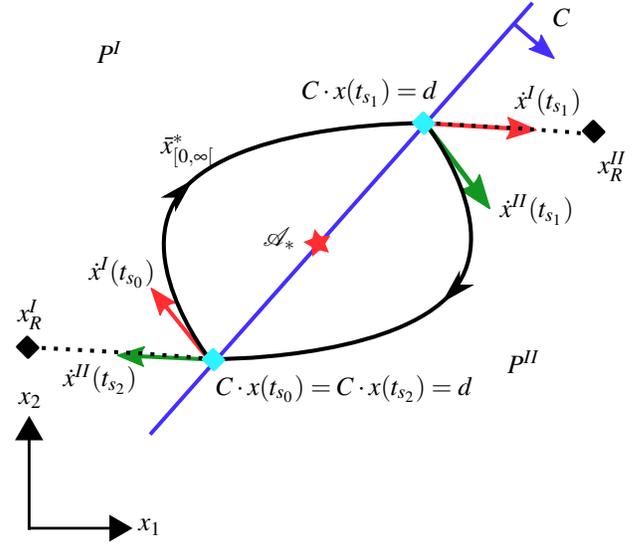}
    \caption{Sufficient conditions \eqref{EQ8}-\eqref{EQ14} for a stable limit cycle $\bar{x}^{*}_{\lbrack0,\infty\lbrack}$.}
    \label{fig:fig1}
\end{figure}

The meaning of \eqref{EQ8}-\eqref{EQ14} for a  limit cycle $\bar{x}^{*}_{\lbrack0,\infty\lbrack}$ of $\Sigma$ is illustrated in Fig. \ref{fig:fig1}, using the abbreviations $\dot{x}^{\RN{1}}(t_{s_k}):=A^{\RN{1}}\cdot x(t_{s_k})+b^{\RN{1}}$, and $\dot{x}^{\RN{2}}(t_{s_k}):=A^{\RN{2}}\cdot x(t_{s_k})+b^{\RN{2}}$ for $k\in\{0,1,2\}$.

\begin{proof}[Thm. 1]
Given $C\cdot x(t_{s_0})=d$ and $C\cdot x(t_{s_1})=d$, any point on the limit cycle $\bar{x}^{*}_{\lbrack0,\infty\lbrack}$ for $t\in[t_{s_0},t_{s_1}]$ follows by integrating \eqref{EQDynI} from:
\begin{equation}
\label{EQ15}
x(t)=e^{A^{\RN{1}}(t_{s_1}-t_{s_0})}\cdot x(t_{s_0})+\int\limits_{t_{s_0}}^{t_{s_1}} e^{A^{\RN{1}}(t_{s_1}-\tau)}\cdot b^{\RN{1}}\,\mathrm{d}\tau,
\end{equation}

Due to condition \eqref{EQ8}, $\Sigma$ is forced to activate the dynamics \eqref{EQDynI} during $t\in[t_{s_0},t_{s_1}]$ with $x(t)\in P^{\RN{1}}$. Since \eqref{EQDynI} is stable according to the assumptions on $A^{\RN{1}}$, the state is attracted to $x_R^{\RN{1}}$, which is, however, positioned in 
$P^{\RN{2}}$ according to the first part of condition \eqref{EQ10}. This together with the  condition \eqref{EQ9} enforces the first switching event at a finite time $t_{s_1}$, by which $x(t)$ transitions from $P^{\RN{1}}$ to $P^{\RN{2}}$, and the dynamics \eqref{EQDynII} is activated for $t\in [t_{s_1},t_{s_2}]$. With $C\cdot x(t_{s_2})=d$ and by integrating \eqref{EQDynII}, any point on  
$\bar{x}^{*}_{\lbrack0,\infty\lbrack}$ for $t\in[t_{s_1},t_{s_2}]$ is given by:
\begin{equation}
\label{EQ16}
x(t)=e^{A^{\RN{2}}(t_{s_2}-t_{s_1})}\cdot x(t_{s_1})+\int\limits_{t_{s_1}}^{t_{s_2}} e^{A^{\RN{2}}(t_{s_2}-\tau)}\cdot b^{\RN{2}}\,\mathrm{d}\tau.
\end{equation}
For this phase, the second part of condition \eqref{EQ9} together with the first statement in \eqref{EQ14} and \eqref{EQ11} implies that $x(t)$ is governed by the stable dynamics \eqref{EQDynII} and is attracted to the stable equilibrium point $x_R^{\RN{2}}$, which is contained in $P^{\RN{1}}$ according to the second part of condition \eqref{EQ10}. Thus, the second switching event takes place at a finite time $t_{s_2}$, enforcing that $P^{\RN{2}}$ is left, and the same situation as for the beginning of the first phase holds again. Due to the first part of condition \eqref{EQ14}, the point of initialization $x_{s_0}$ is reached, and the limit cycle is closed by concatenation of the two pieces of $\bar{x}^{*}_{\lbrack0,\infty\lbrack}$ for $[t_{s_0},t_{s_1}]$ and $[t_{s_1},t_{s_2}]$, where for both phases the assumption on the parameters of \eqref{EQDynI} and \eqref{EQDynII} as stated in the theorem ensure unique solutions by \eqref{EQ15} and \eqref{EQ16}. Obviously, the period $T$ of one cycle is equal to the sum of the two phases, as implied by the second part of \eqref{EQ14}. By repeated concatenation of the two alternating phases the complete and unique limit cycle  $\bar{x}^{*}_{\lbrack0,\infty\lbrack}$
is obtained.
\end{proof}

The principle of the conditions stated in Theorem 1 for realizing the limit cycle is that in any phase, the dynamics gears towards an equilibrium point that is unattainable, since a different subsystem is activated before the equilibrium point is reached. At the same time, the gradient conditions formulated for the switching line are determined such that the line definitely crossed before the turn of the trajectory towards the currently relevant equilibrium point occurs. 

Note that the formulation in Theorem 1 with choosing the initial state $x(t_{s_0})$ being positioned on the switching line is used only for notational convenience, i.e., the extension to assigning $x(0)$ to any arbitrary point on the limit cycle is straightforward. 

\begin{figure}[t!]
    \centering
    \psfrag{a}[][]{$C\cdot x=d$}
    \psfrag{a1}[][]{$4$}
    \psfrag{a2}[][]{$2$}
    \psfrag{a3}[][r]{$\Tilde{x}(\Tilde{t}_{s_1})$}
    \psfrag{a4}[][r]{$\Tilde{x}(\Tilde{t}_{s_3})$}
    \psfrag{a5}[][r]{$\Tilde{x}(\Tilde{t}_{s_5})$}
    \psfrag{a6}[][r]{$x(t_{s_0})=x(t_{s_2})$}
    \psfrag{a7}[][l]{$\Tilde{x}(\Tilde{t}_{s_4})$}
    \psfrag{a8}[][l]{$\Tilde{x}(\Tilde{t}_{s_2})$}
    \psfrag{a9}[][]{$1$}
    \psfrag{a10}[][]{$3$}
    \psfrag{a11}[][]{$x(t_{s_1})$}
    \psfrag{a12}[][]{$\Tilde{x}(0)$}
    \psfrag{a13}[][]{}
    \psfrag{a14}[][r]{$\bar{x}^*_{[0,\infty[}$}
    \psfrag{a15}[][]{$\Tilde{x}_{[\Tilde{t}_{s_1},\Tilde{t}_{s_2}]}$}
    \psfrag{a16}[][]{$\Tilde{x}_{[\Tilde{t}_{s_2},\Tilde{t}_{s_3}]}$}
    \psfrag{a17}[][r]{$\Tilde{x}(0)$}
    \psfrag{a18}[][]{$\Tilde{x}(\Tilde{t}_{s_4})$}
    \psfrag{a19}[][]{$\Tilde{x}(\Tilde{t}_{s_2})$}
    \psfrag{a20}[][r]{$\Tilde{x}(\Tilde{t}_{s_1})$}
    \psfrag{a21}[][]{$\Tilde{x}(\Tilde{t}_{s_3})$}
    \psfrag{a22}[][]{$\Tilde{x}(\Tilde{t}_{s_5})$}
    \psfrag{x1}[][]{$P^{\RN{1}}$}
    \psfrag{x2}[][]{$P^{\RN{2}}$}
    \psfrag{i1}[][]{$x_1$}
    \psfrag{i2}[][]{$x_2$}
    \includegraphics[width=0.35\textwidth]{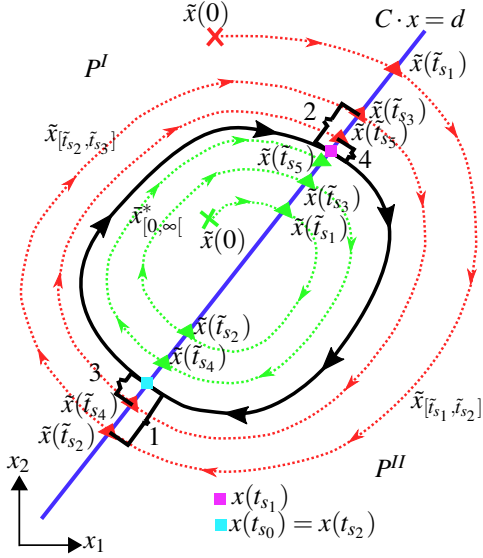}
    \caption{Convergence for an initialization $\Tilde{x}(0)$ in the interior (green) and exterior (red) of $\bar{x}^*_{[0,\infty[}$ (black) with switching line $C\cdot x=d$ (blue).}
    \label{fig:fig1c}
\end{figure}

Next global stability of the limit cycle is considered and discussed based on the Fig. \ref{fig:fig1c} which is oriented to the structure of Fig. \ref{fig:fig1}. In Fig.~\ref{fig:fig1c}, an arbitrary initialization of the state to an $\Tilde{x}(0)$ outside of the limit cycle (black curve) is chosen, and the run from this point is denoted by $\Tilde{x}_{[0,\infty[}$ (shown by a red dashed line). The alternating activation of the two dynamics \eqref{EQDynI} and \eqref{EQDynII} follows the same pattern as explained in the proof of Theorem 1 for the motion on $\bar{x}^*_{[0,\infty[}$: After the first phase of $\Tilde{x}_{[0,\infty[}$ in $P^{\RN{1}}$ with \eqref{EQDynI} for the time interval $t\in[0,\Tilde{t}_{s_1}]$, a switching to \eqref{EQDynII} in $P^{\RN{2}}$ takes place before the first dynamics is again activated at $\Tilde{t}_{s_2}$, and so on. It will be shown in a subsequent theorem that, due to the construction of $\Sigma$ according to Theorem 1, the run $\Tilde{x}_{[0,\infty[}$ approaches $\bar{x}^*_{[0,\infty[}$ in a spiral from the outside. In contrast, Fig. \ref{fig:fig1c} (green dashed line) shows for the initialization to an $\Tilde{x}(0)$ inside of the limit cycle that the run $\Tilde{x}_{[0,\infty[}$ converges to 
$\bar{x}^*_{[0,\infty[}$ from the inside.
To formalize this argumentation, a definition of stability and a corresponding theorem are stated next.

\begin{Def}
\label{Definition3}{Stability of a Limit Cycle of $\Sigma$}\\
A limit cycle $\bar{x}^{*}_{\lbrack0,\infty\lbrack}$ is called \emph{globally stable}, if independent of the initialization $x(0)=x_0\in \mathbb{R}^{2}$ every trajectory converges towards $\bar{x}^{*}_{\lbrack0,\infty\lbrack}$.
 \hfill $\Box$
\end{Def}

\begin{theorem}
\label{theorem2}
For the system $\Sigma$ according to Def.~1, let a unique limit cycle $\bar{x}^*_{[0,\infty[}$ be obtained
by enforcing the conditions stated in Theorem 1. Then, $\bar{x}^*_{[0,\infty[}$ is globally stable according to Def.~3.\hfill $\Box$
\end{theorem}

\begin{proof} For $\Tilde{x}(0)$ outside of $\bar{x}^*_{[0,\infty[}$, convergence of $\Tilde{x}_{[0,\infty[}$ to $\bar{x}^*_{[0,\infty[}$ requires that the sequence of switching points $\Tilde{x}(\Tilde{t}_{s_1}), \Tilde{x}(\Tilde{t}_{s_3}), \Tilde{x}(\Tilde{t}_{s_5}),\ldots$ converges to $x(t_{s_1})$, and likewise that the sequence $\Tilde{x}(\Tilde{t}_{s_2}), \Tilde{x}(\Tilde{t}_{s_4}), \Tilde{x}(\Tilde{t}_{s_5}),\ldots$ converges to $x(t_{s_0})$:
\begin{align}
 &\lim\limits_{i \to \infty} \|\Tilde{x}(\Tilde{t}_{s_i})-x(t_{s_1})\|=0 \text{ for odd } i,\label{EQTH1}\\
 &\lim\limits_{i \to \infty} \|\Tilde{x}(\Tilde{t}_{s_i})-x(t_{s_2})\|=0 \text{ for even } i\label{EQTH2}
\end{align}
with $x(t_{s_2})=x(t_{s_0})$; see also Fig.~\ref{fig:fig1c} (marked in red). The existence of the switching points $\Tilde{x}(t_{s_i})$ for finite $\Tilde{t}_{s_i}$
as reached by continuous evolution from $\tilde{x}(t_{s_{i-1}})$
follows from the stability of $A^{\rm{I}}$ (or $A^{\rm{II}}$ respectively) and the associated equilibrium points $x_R^{\rm{I}}$ (or $x_R^{\rm{II}}$) on the reverse side of the switching line. The same reasoning leads to the existence of $\Tilde{x}(t_{s_{1}})$ as reached from $\Tilde{x}(0)$.

In order to establish \eqref{EQTH1} and 
\eqref{EQTH2}, the mapping from $\Tilde{x}(\Tilde{t}_{s_i})$ to the next intersection with the switching line in the same direction is: 
\begin{align}
\label{EQPTH2}
\begin{split}
\Tilde{x}(\Tilde{t}_{s_{i+2}})=e^{A^{\RN{1}}(\Tilde{t}_{s_{i+2}}-\Tilde{t}_{s_{i+1}})}e^{A^{\RN{2}}(\Tilde{t}_{s_{i+1}}-\Tilde{t}_{s_{i}})}\Tilde{x}(\Tilde{t}_{s_i})\\
+e^{A^{\RN{1}}(\Tilde{t}_{s_{i+2}}-\Tilde{t}_{s_{i+1}})}\int\limits_{\Tilde{t}_{s_i}}^{\Tilde{t}_{s_{i+1}}}e^{A^{\RN{2}}(\Tilde{t}_{s_{i+1}}-\tau)}b^{\RN{2}} d\tau+\int\limits_{\Tilde{t}_{s_{i+1}}}^{\Tilde{t}_{s_{i+2}}}e^{A^{\RN{1}}(\Tilde{t}_{s_{i+2}}-\tau)}b^{\RN{1}} d\tau
\end{split}
\end{align}
The sequence of distances of the switching points $\Tilde{x}(\Tilde{t}_{s_i})$ for odd $i$ to $x(t_{s_1})$ (and of $\Tilde{x}(\Tilde{t}_{s_i})$ for even $i$ to $x(t_{s_2})$ respectively) is measured by a discrete Lyapunov function, defined only on the switching line ($C\cdot\Tilde{x}(\Tilde{t}_{s_i})=d$) with $\Delta x_d=\Tilde{x}(\Tilde{t}_{s_i})-x(t_{s_1})$ for odd $i$, and $\Delta x_d=\Tilde{x}(\Tilde{t}_{s_i})-x(t_{s_2})$ for even $i$.
\begin{align}
V_d(\Tilde{x}(\Tilde{t}_{s_i}))=\|\Delta x_d\|^2=\Delta x_d^T\Delta x_d,~ d\in \{1,2\}.
\end{align}
Similar to the procedure in \cite{Rubensson.1998,Peleties.1991}, decrease of $V_d$ according to:
\begin{align}
\label{EQPTH3}
V_{d}(\Tilde{x}(\Tilde{t}_{s_{i+2}}))-V_{d}(\Tilde{x}(\Tilde{t}_{s_i}))<0
\end{align}
results as follows: 


Define the convergence ratio $\alpha_{\Tilde{t}_{s_{i+2}},\Tilde{t}_{s_i}}$ for a $2$-periodic limit cycle gives a recursive definition of the distances marked by 1 and 3, respectively 2 and 4 in Fig. \ref{fig:fig1c}:
\begin{align}
\label{EQPTH4}
\frac{\|\Tilde{x}(\Tilde{t}_{s_{i+2}})-x(t_{s_d})\|^2}{\|\Tilde{x}(\Tilde{t}_{s_i})-x(t_{s_d})\|^2}\leq \alpha_{\Tilde{t}_{s_{i+2}},\Tilde{t}_{s_i}}
\end{align}
Thus, \eqref{EQPTH4} and the definition of the discrete Lyapunov functions in \eqref{EQPTH3} give a relation for any switch:
\begin{align}
\label{EQPTH5}
V_{d}(\Tilde{x}(\Tilde{t}_{s_{i+2}}))\leq \alpha_{\Tilde{t}_{s_{i+2}},\Tilde{t}_{s_i}} V_{d}(\Tilde{x}(\Tilde{t}_{s_i}))
\end{align}
Given a stable limit cycle with switching sequence $\Tilde{t}_{s_1}, \Tilde{t}_{s_2}\dots$ then for any initial switching point the solution will asymptotically converge to the limit cycle if:
\begin{align}
\label{EQPTH6}
\alpha_{\Tilde{t}_{s_{i+2}},\Tilde{t}_{s_i}}<1
\end{align}
The recursive representation in \eqref{EQPTH2} can be used for odd and even $i$ to represent the inequality in \eqref{EQPTH4}. Since both subsystems are Hurwitz and the time differences are positive every exponential function can become upper bounded lower than one. By comparing the time differences to an equal representation of the limit cycle (fit \eqref{EQ16} into \eqref{EQ15}) the affine parts are ensured to be lower than one additionally. Thus \eqref{EQPTH4} holds with respect to Theorem 1. The discrete Lyapunov functions decrease and therefore the initialisation $\Tilde{x}(\Tilde{t}_{s_i})$ converges to $x(t_{s_1})$ or $x(t_{s_2})$ on the limit cycle. Finally, the argumentation in the proof of Theorem 1 shows that any initialization to $\Tilde{x}(0)$ on $\bar{x}^*_{[0,\infty[}$ means that $\bar{x}^*_{[0,\infty[}$ is never left, thus completing the proof. The same reasoning can be applied for an initialization $\tilde{x}(0)$ inside of $\bar{x}^*_{[0,\infty[}$ and the situation shown in Fig.~\ref{fig:fig1c} (green), leading to the result that the $\Tilde{x}(\Tilde{t}_{s_i})$ converge from the interior of $\bar{x}^*_{[0,\infty[}$ to $x(t_{s_0})$, or $x(t_{s_1})$ respectively.\\
The initialisation excludes cases in which $\dot{x}=0$ shows up.
\end{proof}

\section{LIMIT CYCLE DESIGN}
Based on the conditions for the existence of limit cycles presented before, this section shows how a specific stable limit cycle can be constructed starting from specifications for the amplitude and frequency of the oscillation of $\Sigma$.\\

\begin{Def}
\label{Definition4}{Amplitude of the Limit cycle}\\
Let $x(t)=[x_1(t),x_2(t)]^T$ describe a point moving along the limit cycle $\bar{x}^{*}_{\lbrack0,\infty\lbrack}$ with period $T$ for $t\in[i\cdot T, (i+1)\cdot T]$, $i\in\mathbb{N}\cup\{0\}$. In addition, let $\mathcal{A}_{*}=[\mathcal{A}^{*}_1,\mathcal{A}^{*}_2]^T$ 
denote the center point in between the two switching points on the switching line $C\cdot x(t_{s_0})=d$ and $C\cdot x(t_{s_1})=d$, see Fig. \ref{fig:fig2}. The time-varying  amplitude $\mathcal{A}(t)$ is then defined as the two  Euclidean norms $\mathcal{A}^{\RN{1}}(t)=\|x(t)-\mathcal{A}_{*}\|_2$ for $x(t)\in P^{\RN{1}}$ and $\mathcal{A}^{\RN{2}}(t)=\|x(t)-\mathcal{A}_{*}\|_2$ for $x(t)\in P^{\RN{2}}$.
Furthermore, let $\mathcal{A}_{max}^{\RN{1}}$ denote the maximum of
$\mathcal{A}^{\RN{1}}(t)=\|x(t)-\mathcal{A}_{*}\|_2$ over $x(t)\in P^{\RN{1}}$, and $\mathcal{A}_{max}^{\RN{2}}$ the maximum of $\mathcal{A}^{\RN{2}}(t)=\|x(t)-\mathcal{A}_{*}\|_2$ over $x(t)\in P^{\RN{2}}$. Finally, $\mathcal{A}_{min}^{\RN{1}}$ and $\mathcal{A}_{min}^{\RN{2}}$ denote the corresponding minimum amplitudes.
\end{Def}\vspace{1mm}

\begin{Def}{Limit cycle frequency}\\
\label{Definition5}
Given the limit cycle $\bar{x}^{*}_{\lbrack0,\infty\lbrack}$ of $\Sigma$ with period $T$ as defined in \eqref{EQ14}, the frequency of the corresponding oscillation of $x(t)$ is $\omega=\frac{2\pi}{T}$.
\end{Def}\vspace{1mm}

To fully determine $\Sigma$ (while considering the conditions provided in Theorem 1), the equations \eqref{EQ15} and \eqref{EQ16} are solved by use of the decomposition:
\newenvironment{salign}{\thinmuskip=1mu\medmuskip=1mu\thickmuskip=1mu\align}{\endalign}
\begin{salign}
\label{EQ17}
e^{A_i\cdot t}=W_i\cdot e^{A_{i_D}\cdot t}\cdot W_i^{-1}=W_i\cdot \left[ \begin{array}{rrr}
e^{\lambda_{1,2}^{i}\cdot t}  & 0  \\ 
0 & e^{\lambda_{2,2}^{i}\cdot t}  \\
\end{array}\right]\cdot W_i^{-1},
\end{salign}
where $A_{i_D} \in\mathbb{R}^{2x2}$ denotes the 
diagonalised matrix to $A_i \in\mathbb{R}^{2x2}$ with eigenvalues 
$\lambda_i=\lbrack \lambda_{1,2}^{i},\lambda_{2,2}^{i} \rbrack^T \in\mathbb{R}$, and the eigenvector matrix $W_i \in\mathbb{R}^{2x2}$.

Solving equation \eqref{EQ15} with \eqref{EQ17} leads to \eqref{EQ18}, while solving \eqref{EQ16} likewise leads to \eqref{EQ19}:
\newcommand{\mysmallarraydecl}{\renewcommand{%
\IEEEeqnarraymathstyle}{\scriptscriptstyle}%
\renewcommand{\IEEEeqnarraytextstyle}{\scriptsize}%
\renewcommand{\baselinestretch}{1}%
\settowidth{\normalbaselineskip}{\scriptsize
\hspace{\baselinestretch\baselineskip}}%
\setlength{\baselineskip}{\normalbaselineskip}%
\setlength{\jot}{0.25\normalbaselineskip}%
\setlength{\arraycolsep}{2pt}}
\let\originalleft\left
\let\originalright\right
\renewcommand{\left}{\mathopen{}\mathclose\bgroup\originalleft}
\renewcommand{\right}{\aftergroup\egroup\originalright}
\begin{salign}
\label{EQ18}
x(t) = \left[\begin{IEEEeqnarraybox*}[][c]{,c,}
e^{\lambda^{\RN{1}}_{1,1}t_{d_1}}(S_1+\frac{S_5}{\lambda^{\RN{1}}_{1,1}}) +e^{\lambda^{\RN{1}}_{2,1}t_{d_1}}(S_2+\frac{S_6}{\lambda^{\RN{1}}_{2,1}})+H_1\\
e^{\lambda^{\RN{1}}_{1,1}t_{d_1}}(S_3+\frac{S_7}{\lambda^{\RN{1}}_{1,1}}) + e^{\lambda^{\RN{1}}_{2,1}t_{d_1}}(S_4+\frac{S_8}{\lambda^{\RN{1}}_{2,1}})+H_2
\end{IEEEeqnarraybox*}\right]
\end{salign}
\begin{salign}
\label{EQ19}
x(t) = \left[\begin{IEEEeqnarraybox*}[][c]{,c,}
e^{\lambda^{\RN{2}}_{1,2}t_{d_2}}(S_9+\frac{S_{13}}{\lambda^{\RN{2}}_{1,2}}) +e^{\lambda^{\RN{2}}_{2,2}t_{d_2}}(S_{10}+\frac{S_{14}}{\lambda^{\RN{2}}_{2,2}})+H_3\\
e^{\lambda^{\RN{2}}_{1,2}t_{d_2}}(S_{11}+\frac{S_{15}}{\lambda^{\RN{2}}_{1,2}}) + e^{\lambda^{\RN{2}}_{2,2} t_{d_2}}(S_{12}+\frac{S_{16}}{\lambda^{\RN{2}}_{2,2}})+H_4
\end{IEEEeqnarraybox*}\right]
\end{salign}
\begin{salign}
\label{EQ19A}
H_1=-\frac{S_5}{\lambda^{\RN{1}}_{1,1}}-\frac{S_6}{\lambda^{\RN{1}}_{2,1}},\enspace H_2=-\frac{S_7}{\lambda^{\RN{1}}_{1,1}}-\frac{S_8}{\lambda^{\RN{1}}_{2,1}},~t_{d_1}=t_{s_1}-t_{s_0}\\
 H_3=-\frac{S_{13}}{\lambda^{\RN{2}}_{1,2}}-\frac{S_{14}}{\lambda^{\RN{2}}_{2,2}},\enspace H_4=-\frac{S_{15}}{\lambda^{\RN{2}}_{1,2}}-\frac{S_{16}}{\lambda^{\RN{2}}_{2,2}},~t_{d_2}=t_{s_2}-t_{s_1}
\end{salign}
The substitutions $S_1$ to $S_{16}$ used in these equations are listed in the appendix.

To determine a limit cycle complying with the specifications, the entries of $A^{\RN{1}}$, $b^{\RN{1}}$, $A^{\RN{2}}$ and $b^{\RN{2}}$ are computed. Since each subsystem has six degrees of freedom, a total of twelve equations is needed to obtain a stable limit cycle by two switching planar affine subsystems. Selected points are used to fix the degrees of freedom. Two of these points are the switching points satisfying $C\cdot x(t_{s_1})=d$ and $C\cdot x(t_{s_2})=d$, used twice for both subsystems. (Inserting the coordinates of these points into \eqref{EQ18} and \eqref{EQ19} leads to 
eight equations.) Furthermore, for each of the two polytopes $P^{\RN{1}}$ and $P^{\RN{2}}$, an arbitrarily chosen additional point can be selected. Using a maximum or minimum amplitude appears as reasonable choice to determine these points. If these are inserted into  
\eqref{EQ18} and \eqref{EQ19}, the required set of twelve equations is obtained to fix all degrees of freedom. These equations together with the conditions of Theorem \ref{Theorem1} determine $\Sigma$ as well as the limit cycle.

\subsection{Example}
\begin{figure}[h!]
    \centering
    \psfrag{a}[][]{$x_1$}
    \psfrag{a1}[][]{$x_2$}
    \psfrag{a2}[][]{-2.5}
    \psfrag{a3}[][]{-4}
    \psfrag{a4}[][]{-5.5}
    \psfrag{a5}[][]{-0.25}
    \psfrag{a6}[][]{0}
    \psfrag{a7}[][]{0.25}
    \psfrag{a8}[][]{$C\cdot x=d$}
    \psfrag{a9}[][]{$x^{\RN{1}}_{min}(t_{min}^{\RN{1}})$}
    \psfrag{a10}[][r]{$\mathcal{A}^{\RN{1}}_{min}(t_{min}^{\RN{1}})$}
    \psfrag{a11}[][]{$x^{\RN{2}}_{max}(t_{max}^{\RN{2}})$}
    \psfrag{a12}[][r]{$\mathcal{A}^{\RN{2}}_{max}(t_{max}^{\RN{2}})$}
    \psfrag{a13}[][]{$x_{sp}(t_{s_0}),x_{sp}(t_{s_1})$}
    \psfrag{a14}[][]{$\mathcal{A}_*$}
    \includegraphics[width=0.3\textwidth]{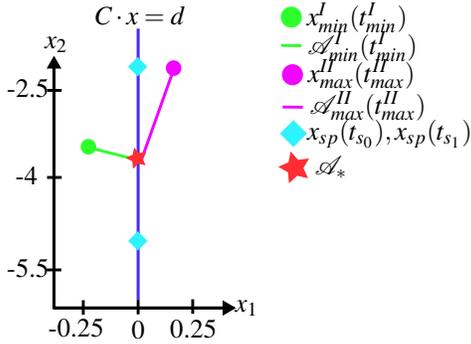}
    \caption{Characterizing points of the design example.}
    \label{fig:fig2}
\end{figure}

\begin{figure}[h!]
\centering
\psfrag{-2}[][]{\footnotesize $-2$}
\psfrag{-2.5}[][]{\footnotesize$-2.5$}
\psfrag{-3}[][]{\footnotesize$-3$}
\psfrag{-3.5}[][]{\footnotesize$-3.5$}
\psfrag{-4}[][]{\footnotesize$-4$}
\psfrag{-4.5}[][]{\footnotesize$-4.5$}
\psfrag{-5}[][]{\footnotesize$-5$}
\psfrag{-1.5}[][]{\footnotesize $-1.5$}
\psfrag{-1}[][]{\footnotesize $-1$}
\psfrag{-0.5}[][]{\footnotesize $-0.5$}
\psfrag{0}[][]{\footnotesize $0$}
\psfrag{0.5}[][]{\footnotesize $0.5$}
\psfrag{1}[][]{\footnotesize $1$}
\psfrag{1.5}[][]{\footnotesize $1.5$}
\psfrag{2}[][]{\footnotesize $2$}
\psfrag{a}[][]{$x_1$}
\psfrag{a1}[][]{$x_2$}
\psfrag{a2}[][r]{$x(0)$}
    \includegraphics[width=0.35\textwidth]{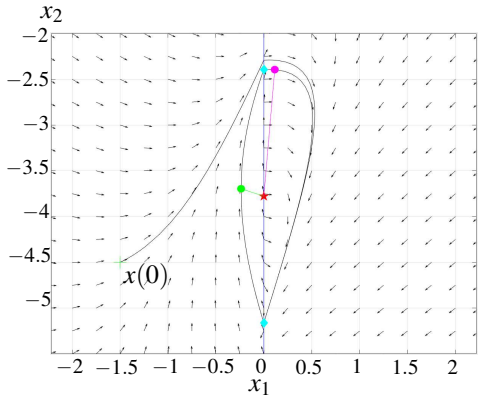}
    \caption{Limit cycle with characterizing points of the design example.}
    \label{fig:fig3}
\end{figure}

\begin{figure}[t!]
\centering
\psfrag{0}[][]{\footnotesize $0$}
\psfrag{5}[][]{\footnotesize$5$}
\psfrag{10}[][]{\footnotesize$10$}
\psfrag{0.2}[][]{\footnotesize$0.2$}
\psfrag{0.4}[][]{\footnotesize$0.4$}
\psfrag{0.6}[][]{\footnotesize$0.6$}
\psfrag{0.8}[][]{\footnotesize$0.8$}
\psfrag{1}[][]{\footnotesize $1$}
\psfrag{1.2}[][]{\footnotesize $1.2$}
\psfrag{1.4}[][]{\footnotesize $1.4$}
\psfrag{1.6}[][]{\footnotesize $1.6$}
\psfrag{a}[][]{$t$}
\psfrag{a1}[][]{$\mathcal{A}(t)$}
\psfrag{a2}[][]{$\mathcal{A}^{\RN{2}}_{max}(t_{max}^{\RN{2}})$}
\psfrag{a3}[][r]{$\mathcal{A}^{\RN{1}}_{min}(t_{min}^{\RN{1}})$}
    \includegraphics[width=0.25\textwidth]{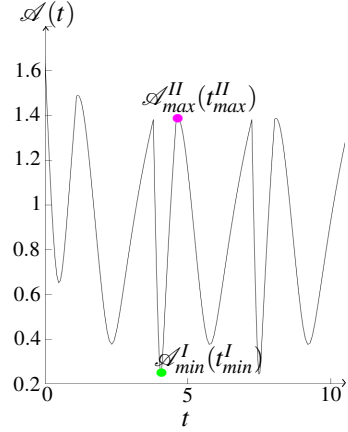}
    \caption{Amplitude of the limit cycle for the design example.}
    \label{fig:fig4}
\end{figure}

The aforementioned procedure is illustrated by an example: Assume  that the frequency  $\omega=1.824$Hz of a limit cycle, the minimum amplitude of $\mathcal{A}^{\RN{1}}_{min}(t_{min}^{\RN{1}})=0.25$, and the maximum amplitude $\mathcal{A}^{\RN{2}}_{max}(t_{max}^{\RN{2}})=1.3778$ (for $t_{min}^{\RN{1}},t_{max}^{\RN{2}}\in \mathbb{R}_{\geq 0}$) are given as specifications for system design. 
The switching line is defined by $C=[1, 0]$ and $d=0$. Since $T$ is equal to the sum of the two phases, $t_{s_1}=0.801$s and $t_{s_2}=2.644$s are chosen, leading to switching points $x_{sp}(t_{s_1})=[0,-2.394]^T$ and $x_{sp}(t_{s_0})=x_{sp}(t_{s_2})=[0,-5.160]^T$. Using $\mathcal{A}_*=[0,-3.778]^T$ and Def.~4 results in $x^{\RN{1}}_{min}(t_{min}^{\RN{1}})=[-0.237,-3.6959]^T$, $t_{min}^{\RN{1}}=0.2825$s, $t_{min}^{\RN{1}}\leq t_{s_1}$ and $x^{\RN{2}}_{max}(t_{max}^{\RN{2}})=[0.1142,-2.395]^T$, as well as $t_{max}^{\RN{2}}=0.0482$s, $t_{max}^{\RN{2}}\leq t_{s_2}$. Relevant points of construction for this example are illustrated in Fig.\ref{fig:fig2}.
The four points and their coordinates determine twelve equations as previously explained. Solving the equation system leads to an oscillator system of type  $\Sigma$ with two affine dynamics parameterized by:
\begin{equation}
\label{EQ21}
A^{\RN{1}}=\begin{bmatrix}
-3&1 \\
3&-2 \\
\end{bmatrix}, \enspace b^{\RN{1}}=\begin{bmatrix}
3 \\
-3 \\
\end{bmatrix}
\end{equation}
\begin{equation}
\label{EQ22}
A^{\RN{2}}=\begin{bmatrix}
-4&1 \\
-3&0.25 \\
\end{bmatrix}, \enspace b^{\RN{2}}=\begin{bmatrix}
5 \\
0.75 \\
\end{bmatrix}
\end{equation}
The Figs. \ref{fig:fig3} and  \ref{fig:fig4} show the resulting stable limit cycle and the course of the  amplitude over time, starting from an arbitrarily chosen initial state $x(0)=[-1.5,-4.5]^T$. The  characterizing points used for design are, of course, located on the limit cycle.


\section{LIMIT CYCLE IDENTIFICATION}
\begin{figure}[h!]
\centering
\psfrag{0.3}[][]{\footnotesize $0.3$}
\psfrag{0.4}[][]{\footnotesize$0.4$}
\psfrag{0.5}[][]{\footnotesize$0.5$}
\psfrag{0.6}[][]{\footnotesize$0.6$}
\psfrag{0.7}[][]{\footnotesize$0.7$}
\psfrag{0.8}[][]{\footnotesize$0.8$}
\psfrag{0.9}[][]{\footnotesize$0.9$}
\psfrag{1}[][]{\footnotesize $1$}
\psfrag{1.1}[][]{\footnotesize $1.1$}
\psfrag{1.2}[][]{\footnotesize $1.2$}
\psfrag{a}[][]{$C,d$}
\psfrag{a1}[][]{\eqref{EQ25}}
\psfrag{a2}[][r]{$\Sigma_{Pd}$}
\psfrag{a3}[][r]{$O_{Pd}$}
\psfrag{a4}[][]{$x(0)$}
\psfrag{a5}[][]{$y_2,x_2$}
\psfrag{a6}[][]{$y_1,x_1$}
    \includegraphics[width=0.33\textwidth]{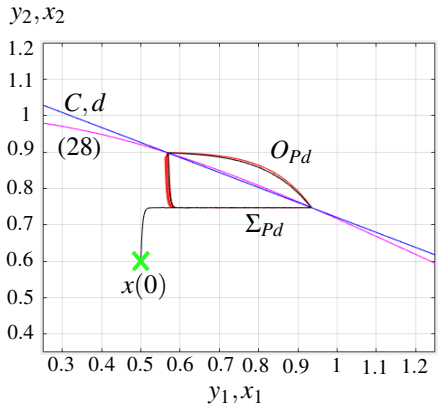}
    \caption{Identification of the limit cycle for the reaction system.}
    \label{fig:fig5}
\end{figure}

By employing the design principles introduced before, this section proposes  an approach to identify stable limit cycles from measured (possibly noisy) data of nonlinear systems. Consider an oscillating chemical reaction, a so-called \emph{relaxation oscillator}, which is characterized by alternating phases of fast and slow reaction. The oscillator shows the basic mechanism of cyclical oxidation and reduction of palladium. Let $y_1$ describe the  state of oxidation of a palladium catalyst and $y_2$ the CO concentration in the reactor. Additionally, $\alpha$ denotes the flow rate through the reactor, and $Q$ indicates the division into active and passive regions. For the purpose of this section, the following model is subsequently used for data generation -- in practice, however, the model would be supposed to be unknown, and only periodic data would be available:
\begin{salign}
    \label{EQ23}
    \dot{y_1}=[\Theta(y_1,y_2,Q)-y_1]\cdot \beta
\end{salign}
\begin{salign}
    \label{EQ24}
    \dot{y_2}=-\Theta(y_1,y_2,Q)\cdot y_2+\alpha\cdot y_0-\alpha\cdot[1-\Theta(y_1,y_2,Q)]\cdot y_2
\end{salign}
\begin{salign}
    \label{EQ25}
    \Theta(y_1,y_2,Q)=\Theta_0\cdot (f(y_1,Q)-y_2),~f(y_1,Q)=e^\frac{-y_1^2}{Q}
\end{salign}
 $\Theta_0$ is the Heaviside step function, and the frequency $\beta$ of the oscillation is:
\begin{salign}
    \label{EQ26}
    \beta=\Theta(y_1,y_2,Q)\cdot \overline{\beta}+(1-\Theta(y_1,y_2,Q))\cdot \beta_0
\end{salign}
with $\beta_0$ describing the speed of reduction, and $\overline{\beta}$ the speed of oxidation. The $y_1$-$y_2$ phase space is divided by the boundary line $f(y_1,Q)$ (see Fig. \ref{fig:fig5} magenta) into an active and a passive area with different dynamic behaviour. In the active area, equation \eqref{EQ25} is used in \eqref{EQ23}, \eqref{EQ24} and \eqref{EQ26} being $1$ and $0$ in the passive area.
White noise is added to the right-hand sides of \eqref{EQ23} and \eqref{EQ24} to represent unknown influences of the chemical process and generate appropriated data for the identification process. More detailed information on the model can be found in \cite{Ballandis.2001}.

Given data from the simulation of the model (for a flow rate of $\alpha=0.83$), the objective is to identify a PSAS that represents the oscillating behavior. When detecting the switching points as well as two more characteristic points, the aforementioned procedure can be used to obtain the model $\Sigma$ with the following parameterization of the affine dynamics:
\begin{equation}
\label{EQ27}
A^{\RN{1}}=\begin{bmatrix}
-0.01\cdot \alpha &0 \\
0&-1 \\
\end{bmatrix}, \enspace b^{\RN{1}}=\begin{bmatrix}
0.01\cdot \alpha \\
0.9\cdot \alpha \\
\end{bmatrix}
\end{equation}
\begin{equation}
\label{EQ28}
A^{\RN{2}}=\begin{bmatrix}
0.1&0 \\
0&-\alpha \\
\end{bmatrix}, \enspace b^{\RN{2}}=\begin{bmatrix}
0 \\
0.9\cdot \alpha \\
\end{bmatrix}
\end{equation}
The nonlinear switching line (Fig. \ref{fig:fig5} magenta) was approximated by $C=[0.4115, 1]$ and $d=1.132$ (Fig. \ref{fig:fig5} blue). This line was determined based on the two switching points which divide the active and passive regions. For the identification, the maxima of the amplitudes on the two regions were used to generate the corresponding equations.
Thus, the scheme as in Fig. \ref{fig:fig2} can be applied and enables successful identification. Fig. \ref{fig:fig5} compares the limit cycle from the model for data generation (referred to by $O_{Pd}$) for $\alpha=0.83$ in red (without noise), and the limit cycle of the model identified as PSAS (referred to by $\Sigma_{Pd}$) in black. The difference is negligible, i.e., the model $\Sigma_{Pd}$ (whose structure is significantly simple than that of $O_{Pd}$) can be used for the analysis of the system.
\section{CONCLUSIONS}
 The paper has proposed a new method for synthesizing models of type planar switching affine systems to represent oscillating behavior. The design rules
guarantee the global stability and uniqueness of the resulting limit cycle. The advantages of the model are manifold: (i) linear system theory is sufficient to analyze the oscillating behavior; (ii) the design of affine controllers for the two affine dynamics is suited to instantiate the embedded subsystems; (iii) the oscillations are robust in the sense that (due to the property of global stability) deviations from the limit cycles do not lead to divergence from $\bar{x}^*_{[0,\infty[}$;
(iv) the design rules build on very few specifications for amplitudes and the frequency to obtain the desired oscillations -- in system identification, the low effort for design and the few parameters in the model have to be contrasted to numeric procedures (such as machine learning) to obtain a model with typical many parameters from a large set of data.

Future work includes investigation in higher dimensions and extensions to more subsystems and switching surfaces. 
In addition,
the coupling of several oscillators of the proposed type will be investigated, as well as the application for biological systems with respect to modeling periodic rhythms.  
%




\section*{APPENDIX}
\subsection{Substitutions used for computing the limit cycle analytically: If the index of a Substitution is lower or equal than eight choose $R:=\RN{1}$, $e:=0$ otherwise choose $R:=\RN{2}$, $e:=1$.}
\begin{salign}
    S_{1,9}=\begin{IEEEeqnarraybox*}[\mysmallarraydecl][c]{,c,}
\frac{x_1(t_{s_e})v^{R}_{1,1}v^{R}_{2,2}-x_2(t_{s_e})v^{R}_{1,1}v^{R}_{1,2}}{v^{R}_{1,1}v^{R}_{2,2}-v^{R}_{2,1}v^{R}_{1,2}},\enspace
\end{IEEEeqnarraybox*}
S_{2,10}=\begin{IEEEeqnarraybox*}[\mysmallarraydecl][c]{,c,}
\frac{-x_1(t_{s_e})v^{R}_{1,2}v^{R}_{2,1}+x_2(t_{s_e})v^{R}_{1,1}v^{R}_{1,2}}{v^{R}_{1,1}v^{R}_{2,2}-v^{R}_{2,1}v^{R}_{1,2}}
\end{IEEEeqnarraybox*}
\end{salign}
\begin{salign}
    S_{3,11}=\begin{IEEEeqnarraybox*}[\mysmallarraydecl][c]{,c,}
\frac{x_1(t_{s_e})v^{R}_{2,1}v^{R}_{2,2}-x_2(t_{s_e})v^{R}_{2,1}v^{R}_{1,2}}{v^{R}_{1,1}v^{R}_{2,2}-v^{R}_{2,1}v^{R}_{1,2}},\enspace
\end{IEEEeqnarraybox*}
S_{4,12}=\begin{IEEEeqnarraybox*}[\mysmallarraydecl][c]{,c,}
\frac{-x_1(t_{s_e})v^{R}_{2,2}v^{R}_{2,1}+x_2(t_{s_e})v^{R}_{2,2}v^{R}_{1,1}}{v^{R}_{1,1}v^{R}_{2,2}-v^{R}_{2,1}v^{R}_{1,2}}
\end{IEEEeqnarraybox*}
\end{salign}

\begin{salign}
    S_{5,13}=\begin{IEEEeqnarraybox*}[\mysmallarraydecl][c]{,c,}
\frac{b^{R}_1v^{R}_{1,1}v^{R}_{2,2}-b^{R}_2v^{R}_{1,1}v^{R}_{1,2}}{v^{R}_{1,1}v^{R}_{2,2}-v^{R}_{2,1}v^{R}_{1,2}},\enspace
\end{IEEEeqnarraybox*}
S_{6,14}=\begin{IEEEeqnarraybox*}[\mysmallarraydecl][c]{,c,}
\frac{-b^{R}_1v^{R}_{1,2}v^{R}_{2,1}+b^{R}_2v^{R}_{1,1}v^{R}_{1,2}}{v^{R}_{1,1}v^{R}_{2,2}-v^{R}_{2,1}v^{R}_{1,2}}
\end{IEEEeqnarraybox*}
\end{salign}
\begin{salign}
    S_{7,15}=\begin{IEEEeqnarraybox*}[\mysmallarraydecl][c]{,c,}
\frac{b^{R}_1v^{R}_{2,1}v^{R}_{2,2}-b^{R}_2v^{R}_{2,1}v^{R}_{1,2}}{v^{R}_{1,1}v^{R}_{2,2}-v^{R}_{2,1}v^{R}_{1,2}},\enspace
\end{IEEEeqnarraybox*}
S_{8,16}=\begin{IEEEeqnarraybox*}[\mysmallarraydecl][c]{,c,}
\frac{-b^{R}_1v^{R}_{2,2}v^{R}_{2,1}+b^{R}_2v^{R}_{2,2}v^{R}_{1,1}}{v^{R}_{1,1}v^{R}_{2,2}-v^{R}_{2,1}v^{R}_{1,2}}
\end{IEEEeqnarraybox*}
\end{salign}
\section*{ACKNOWLEDGMENT}
The authors gratefully acknowledge partial financial support from the German Research Foundation (DFG) through the Research Training Group "Biological Clocks on Multiple Time Scales".

\bibliographystyle{IEEEtran.bst}
\bibliography{IEEEabrv,literature}

\end{document}